\documentstyle[12pt,epsfig]{article} 
\def\beq{\begin{equation}}
\def\eeq{\end{equation}}
\def\ber{\begin{eqnarray}}
\def\eer{\end{eqnarray}}

\def\lsim{\ 
 \lower-1.2pt\vbox{\hbox{\rlap{$<$}\lower5pt\vbox{\hbox{$\sim$}}}}\ }
\def\gsim{\ 
 \lower-1.2pt\vbox{\hbox{\rlap{$>$}\lower5pt\vbox{\hbox{$\sim$}}}}\ }

\def\abs#1{\left| #1 \right|}

\def\x{{\mbox{\boldmath $x$}}}
\def\p{{\mbox{\boldmath $p$}}}
\def\tr{{\rm tr}}

\begin{document}

\title{\hfill IUCAA-43/97 \\ \hfill hep-ph/9706378 \bigskip \\ \bf
Space-time description of neutrino flavour oscillations} 
\author{Yu.~V.~Shtanov \\ 
{\it Inter-University Centre for Astronomy and Astrophysics} \\
{\it Post Bag 4, Ganeshkhind, Pune 411 007, India} \medskip \\ 
{\it and} 
\medskip \\  
{\it Bogolyubov Institute for Theoretical Physics, Kiev 252143, 
Ukraine}$^*$} 

\date{Revised \\ September 9, 1997}

\maketitle                       
 
\bigskip 
 
\begin{abstract} 
  Recently the issue of EPR-like correlations in the mutual 
probability of detecting neutrino together with accompanying 
charged lepton has received a new impetus. In this paper we 
describe this effect using the propagators of the particles 
involved in the Schwinger's parametric integral representation. 
We find this description more simple and more suitable to the 
purpose than the usual momentum-space analysis. We consider 
the cases of monochromatic neutrino source, wave packet source, 
and neutrino creation in a localised space-time region. In the 
latter case we note that the space-time oscillation amplitude 
depends on the values of the neutrino masses, and becomes rather 
small for large relative mass differences (mass hierarchy). We 
obtain the expressions for the oscillation and coherence lengths 
in various circumstances. In the region of overlap our results 
confirm those of Dolgov {\em et al.\@} \cite{DMOS}  
\end{abstract}

\bigskip 

{PACS number(s): 14.60.Pq}

\vfill

\hrule \hfill

$^*$Permanent address. E-mail: shtanov@ap3.gluk.apc.org

\newpage 

\sloppy

\section{Introduction} \label{intro}

 Space-time oscillation of neutrino flavour \cite{BV,KP} is 
considered to be the most promising effect observation of which might 
indirectly establish nonzero neutrino mass. By its very nature it 
requires spatio-temporal description of the processes of neutrino 
creation, propagation and detection, and of the similar processes that 
occur with the accompanying particles. Such a description has been 
performed in \cite{Kayser} and  further developed in \cite{GKLL,Rich} 
without ambiguities that sometimes accompany noncritical use of 
neutrino flavour eigenstates. 

 Recently \cite{SWS,DMOS} the issue of EPR-like correlations 
in the mutual probability of detecting both neutrino and 
an accompanying charged lepton 
has received a new impetus. (It was previously considered in 
\cite{Kayser}.) In order to simplify derivation of the basic 
effects the authors 
of \cite{DMOS} combined together descriptions in the configuration 
space and in the momentum space, of the same relevant processes,  
using simultaneously such mutually exclusive notions as sharp wave 
packets in momentum space and definite 
space-time {\em a posteriori\/} trajectories of particles. 
EPR-like experiments of the same type involving neutral kaon 
and $B$~meson oscillations were  
considered in \cite{KS}. In this paper the authors also adopted a 
simple approach using the action values on the particle classical 
trajectories to evaluate the relevant phase factors in the 
probability amplitude. Although the results obtained in such a 
simplified approach are correct, they also might call for a more 
careful derivation. This will be the aim of the present paper. 

 In this paper we try to analyse the phenomenon in a consistent way 
using the propagators of the particles involved in the Schwinger's
parametric integral representation (see Eq.~(\ref{repr}) below). 
We find this description 
more simple and more suitable to the purpose than the usual one 
which employs propagators in momentum space representation. 
This latter involves rather complicated momentum integrations (see, 
e.g., \cite{GKLL,GS}) and, it seems, frequently obscures the physical 
picture of the phenomenon. Our treatment will be general and will 
contain the analysis of the EPR-like experiments of detecting 
neutrino together with the accompanying charged lepton, as well as 
the standard textbook examples of neutrino flavour space-time 
oscillations. 

 After preliminaries in the following section, in Sec.~\ref{mono} 
we consider the case of a monochromatic neutrino source and the 
probability of mutual detection of the neutrino and 
of the accompanying charged lepton. In Sec.~\ref{pack} the effect 
of wave packet neutrino source is analysed. We obtain the expressions 
for the oscillation and coherence lengths in various circumstances. 
The case of neutrino source in a strongly localised space-time 
region  will 
then be considered in Sec.~\ref{local}. In this case the space-time 
oscillation amplitude depends rather strongly on the values of the 
neutrino masses, and becomes rather small for large relative mass 
differences (neutrino mass hierarchy). We summarise our results in 
Sec.~\ref{sum}. In the Appendix we provide an alternative  
derivation of the probability amplitude for the case of 
monochromatic neutrino source, in order to elucidate the difference
between this case and the case of neutrino source strongly 
localised in space-time.

\section{Preliminaries} \label{prel}

 Throughout this paper we consider a process in which neutrino is 
created together with accompanying charged lepton, and afterwards both 
particles are detected. The charged weak currents $\overline l O_\alpha 
\nu$ are involved in the description of this process, where $O_\alpha = 
\gamma_\alpha (1 + \gamma_5)$, and 
$\gamma_5 = i \gamma_0\gamma_1\gamma_2
\gamma_3$. The amplitude of creation of a \mbox{$l$-$\nu$} pair at 
space-time point $x$ is proportional to $O_\alpha J_S^\alpha(x)$, 
with $J_S^\alpha(x)$ being the source current responsible for this 
process. In the case of pion we would have 
$J_S^\alpha(x) \propto \partial^\alpha \phi_\pi(x)$, 
where $\phi_\pi(x)$ is the pion wave function. The charged lepton 
produced at space-time point $x_c$ in a flavour state $a$ can 
propagate to space-time point $x_l$, and the neutrino to space-time 
point $x_n$, at which points these particles may be detected. At the 
space-time point $x_n$ one may detect neutrino-induced charged lepton 
production of flavour $b$.
The amplitude of such a process in which neutrino and the 
corresponding antilepton are created and subsequently detected 
will contain the factor
\beq
O_\beta J_D^\beta \sum_j U_{bj}^\dagger U_{ja} \int dx_c\, 
S_j(x_n,x_c) O_\alpha J_S^\alpha(x_c) S_a(x_c,x_l)\, , \label{ampl}
\eeq
where $U_{ja}$ is the unitary matrix of neutrino mass--flavour mixing 
amplitudes, $U^\dagger_{bj}$ is its Hermitean conjugate, 
$a$ and $b$ numerate flavours, $j$ numerates the neutrino mass 
eigenstates, $S_j$ and $S_a$ are, correspondingly, the Feynman 
propagators of neutrino 
mass specie $j$ with mass $m_j$ and of charged lepton flavour $a$,
and $J_D^\beta$ is the current involved in the neutrino detection 
process, localised around the space-time point $x_n$. 

 The total amplitude that will describe the detection of charged 
lepton and neutrino events will contain, besides the factor 
(\ref{ampl}), also positive and negative energy wave functions
of different finite particles involved in the detection process.
These factors are of particular nature, they do not affect the 
dependence of the amplitude on the space-time coordinates $x_l$ 
and $x_n$, hence they will be omitted as irrelevant to the main 
topic of this paper. Due to these factors, however, as well as, 
in typical cases, due to the positive-frequency 
character of the source, the integration region over $x_c$ in 
(\ref{ampl}) will be effectively restricted to the causal past of 
both points $x_n$ and $x_l$.

 The Feynman propagator $S(x,y) \equiv S_m(x - y)$ for the Dirac 
field of mass $m$ has the form 
\beq
S_m(x) = \left(i\gamma^\alpha\partial_\alpha + m\right) D_m(x)\, , 
\label{prop} 
\eeq
where $D_m(x)$ is the Feynman propagator for the Klein-Gordon field 
of mass $m$. This propagator has the parametric integral 
representation (first considered by Schwinger, Dyson and Feynman in 
the papers collected in \cite{Schwinger})
\beq
D_m(x) = - {1 \over 8\pi^2} \lim_{\epsilon \rightarrow +0} 
\int\limits_0^\infty d \lambda \exp\left[- \frac i2 \left( 
\lambda x^2 + 
\frac 1\lambda \left[m^2 - i\epsilon \right] \right) \right] \, , 
\label{repr}
\eeq
where $x^2 = x \cdot x = x^\alpha x_\alpha$ is the Lorentz interval 
squared. The factors of type $O_\alpha$ in the amplitude 
Eq.~(\ref{ampl}) will have an effect that in the neutrino propagator 
the term proportional to a unit matrix will not contribute, and only 
that proportional to the Dirac gamma matrices will remain. This general 
property is due to the equality $O_\alpha O_\beta = 0$. Thus in 
(\ref{ampl}) we can replace $S_j(x_n,x_c)$ by $\tilde S_j(x_n,x_c) 
\equiv \tilde S_{m_j}(x_n - x_c)$, where
\beq
\tilde S_m(x) = i\gamma^\alpha\partial_\alpha D_{m}(x)\, . 
\label{propn}
\eeq

\section{Monochromatic source} \label{mono}

 In this section we investigate the case of a monochromatic source 
current 
$J_S^\alpha(x)$ that can arise, for instance, in the process of pion 
decay. Let
\beq
J_S^\alpha(x) \propto e^{-i p \cdot x}\, , \label{monc}
\eeq
with constant four-momentum $p$. In the case of pion we would have 
$J_S^\alpha(x) \propto \partial^\alpha \phi_\pi(x) \propto 
p^\alpha \exp(-ip \cdot x)$, where $\phi_\pi(x)$ is the pion wave 
function. We make notation 
\beq
x_{nc} = x_n - x_c \, , \ \ \  x_{lc} = x_l - x_c \, , \ \ \ x_{nl} = 
x_n - x_l\, .
\eeq
In the amplitude (\ref{ampl}) we represent the propagators 
using (\ref{prop})--(\ref{propn}), first perform integration over 
$x_c$, then over the parameters $\lambda_l$ and $\lambda_n$ that 
appear in the representation (\ref{repr}), respectively, for charged 
lepton and neutrino propagators. The integral over $x_c$ is Gaussian,
hence it can be evaluated exactly, pre-exponential factors can be 
obtained after integration over $x_c$ by 
taking partial derivatives with respect to $x_l$ and $x_n$ according 
to (\ref{prop}). The remaining 
integral over $\lambda_l$, $\lambda_n$ will be afterwards evaluated 
in the stationary phase approximation. 

 Consider the integral over $x_c$ of one of the terms in the sum of 
(\ref{ampl}). The phase in the exponent of the integrand will stem 
from the expression (\ref{repr}) for propagators, and from the source 
current in (\ref{ampl}). It will be given by 
\beq
\phi = {} - \frac12 \lambda_l x_{lc}^2 - \frac12 \lambda_n x_{nc}^2 - 
p \cdot x_c \, . \label{phase}
\eeq
Its extremal point $x_c = x_c(\lambda_l, \lambda_n)$ 
is determined from the equation 
\beq
{\partial \phi \over \partial x_c} \equiv \lambda_l x_{lc} + 
\lambda_n x_{nc} - p = 0 \, . \label{extr1}
\eeq
We also have for the matrix of the second derivatives 
\beq
{\partial^2 \phi \over \partial x_c^\alpha \partial x_c^\beta} = 
- (\lambda_l + \lambda_n) g_{\alpha\beta} \, , \label{var1}
\eeq
so that integration over $x_c$ will produce a factor  
\beq
\int dx_c\, e^{i \phi} = {4 i \pi^2 \over \left(\lambda_l + 
\lambda_n\right)^2}\, e^{i \phi_*} \, , \label{fac1}
\eeq
where $\phi_*$ is the value of the phase $\phi$ at the extremal 
point:
\beq
\phi_* = - {\lambda_l \lambda_n x_{nl}^2 - m^2 + 2 p \cdot 
(\lambda_l x_l + \lambda_n x_n) \over 2 
\left(\lambda_l + \lambda_n\right)} \, . \label{phistar}
\eeq 

 Now consider the integral over the $\lambda$'s. It will be 
evaluated in the stationary phase approximation. The phase of 
the integrand is given by 
\beq
\Phi = \phi_* - \frac12 \left({m_l^2 \over \lambda_l} + 
{m_n^2 \over \lambda_n} \right) \, , \label{Phi}
\eeq
where $m_l$ and $m_n$ are the masses, respectively, of charged 
lepton and of neutrino. The stationary point is determined by 
differentiating (\ref{Phi}) using (\ref{phistar}),  
or by the equivalent conditions in the convenient form 
obtained using (\ref{phase}) and (\ref{extr1}), 
\beq
{\partial \Phi \over \partial \lambda_l} \equiv {}- \frac12 x_{lc}^2 
+ {m_l^2 \over 2 \lambda_l^2} = 0\, , \ \ \ 
{\partial \Phi \over \partial \lambda_n} \equiv {}- \frac12 x_{nc}^2 
+ {m_n^2 \over 2 \lambda_n^2} = 0\, . \label{extr2}
\eeq
In these equations $x_c = x_c(\lambda_l, \lambda_n)$ is the solution 
of  Eq.~(\ref{extr1}). From (\ref{extr2}) we have the relation 
\beq
\lambda_l x_{lc} = p_l\, , \ \ \ \lambda_n x_{nc} = p_n \, , 
\eeq
satisfied by the extremal values of $\lambda$'s, 
where $p_l$ and $p_n$ are the four-momenta that the charged lepton and 
the neutrino respectively would have were they free classical particles 
moving from the space-time creation point $x_c$ respectively to the 
registration points $x_l$ and $x_n$.  
Then Eq.~(\ref{extr1}) expresses the energy-momentum conservation law, 
the condition from which the extremal point $x_c$ with extremal 
$\lambda$'s can be found most easily. 
 
 We need also the matrix of the second derivatives of $\Phi$ over 
$\lambda$'s at the extremal point. Differentiating the identity 
(\ref{extr1}) we find
\beq
{\partial x_c \over \partial \lambda_l} = {x_{lc} \over \lambda_l 
+ \lambda_n}\, , \ \ \ 
{\partial x_c \over \partial \lambda_n} = {x_{nc} \over \lambda_l 
+ \lambda_n}\, , \label{partxc}
\eeq
and, differentiating (\ref{extr2}), 
\ber
&{}&{\partial^2 \Phi \over \partial \lambda_l^2} = {}- {x_{lc}^2 
\over \lambda_l + \lambda_n} - {m_l^2 \over \lambda_l^3} = {}- m_l^2\, 
{2 \lambda_l + \lambda_n \over \lambda_l^3 \left( \lambda_l + \lambda_n 
\right)}\, , \label{mat1} \\ 
&{}&{\partial^2 \Phi \over \partial \lambda_n^2} = {} - {x_{nc}^2 
\over \lambda_l + \lambda_n} - {m_n^2 \over \lambda_n^3} = {}- m_n^2\, 
{2 \lambda_n + \lambda_l \over \lambda_n^3 \left( \lambda_l + 
\lambda_n \right)} \, , \label{mat2} \\
&{}&{\partial^2 \Phi \over \partial \lambda_l \partial \lambda_n}
 = {} - {x_{lc} \cdot x_{nc} \over \lambda_l + \lambda_n} = {} - 
{p_l \cdot p_n \over \lambda_l \lambda_n \left(\lambda_l + 
\lambda_n\right)}\, . \label{mat3}
\eer
Let 
\beq
m = \sqrt{p \cdot p}\, , 
\eeq
be the effective mass of the source. In the case of pion decay this 
will be equal to the pion mass $m_\pi$. In a realistic case 
\beq
m - m_l \gg m_n , \ \ \ m_l \gg m_n\, . \label{ms}
\eeq
Below we will see [cf. Eq.~(\ref{lambdas2})] that in the limit 
$m_n \rightarrow 0$ the extremal 
values of $\lambda$'s remain finite. Thus 
we can approximate the determinant of the matrix $\partial^2 \Phi / 
\partial \lambda_i \partial \lambda_j$ by its limit as 
$m_n \rightarrow 0$. The result is 
\beq
\det \left( {\partial^2 \Phi \over \partial \lambda_i \partial 
\lambda_j} \right) \approx - \left({\partial^2 \Phi \over \partial 
\lambda_l \partial \lambda_n} \right)^2 = -
\left({p_l \cdot p_n \over \lambda_l \lambda_n \left(\lambda_l 
+ \lambda_n\right)}\right)^2 \, . \label{det}
\eeq
Therefore the integral over $\lambda$'s will produce a factor 
\beq
{2\pi \lambda_l \lambda_n \left(\lambda_l + \lambda_n \right) \over 
p_l \cdot p_n}\, e^{i\Phi_*} \, , \label{fac2}
\eeq
where $\Phi_*$ is the extremal value of the phase $\Phi$, which is 
given by 
\beq
\Phi_* = {} - m_l \sqrt{x_{lc}^2} - m_n \sqrt{x_{nc}^2} - p 
\cdot x_c\, , \label{Phistar}
\eeq
with $x_c$ being the extremal point of $\phi$ at extremal values of 
$\lambda$'s -- the solution to (\ref{extr1}), (\ref{extr2}). 
Using the extremality conditions (\ref{extr1}), (\ref{extr2}) 
we easily find
\beq
{\partial \Phi_* \over \partial x_l} = - p_l\, , \ \ \ 
{\partial \Phi_* \over \partial x_n} = - p_n\, ,
\eeq
and 
\beq
\Phi_* = {} - p_l \cdot x_l - p_n \cdot x_n = {} - p \cdot x_l - p_n 
\cdot x_{nl} \, . \label{Phi*}
\eeq
Note that the four-momenta $p_l$ and $p_n$ lie in the plane formed 
by the four-vectors $p$ and $x_{nl}$, and are determined by 
energy-momentum conservation. 

 Combining together the factors calculated (\ref{fac1}) and 
(\ref{fac2}), dropping the resulting overall 
numerical constant $i / 8\pi$, and using 
(\ref{Phi*}) we obtain the expression for 
the amplitude (\ref{ampl}) in the case of monochromatic source 
\beq
e^{- i\, p \cdot x_l}\,  \sum_j {\lambda_l \lambda_n 
\over \left(\lambda_l + \lambda_n\right) p_l \cdot p_n}\, 
O_D\gamma_\alpha p_n^\alpha O_S \left(m_l - 
\gamma_\beta p_l^\beta\right)U_{bj}^\dagger U_{ja}\, e^{-i\, p_n \cdot 
x_{nl}} \, , \label{amplmid}
\eeq
where $O_D = O_\alpha J_D^\alpha$, $O_S = O_\alpha J_S^\alpha$. Note 
that the coordinate-dependence of the current $J_S(x)$ has transformed 
to the phase of (\ref{amplmid}). Also note that the extremal values of 
$\lambda$'s as well as the four-momenta $p_l$ and $p_n$ under the sum 
(\ref{amplmid}) depend on the neutrino specie $j$. However, the 
prefactors in our expression (\ref{amplmid}), as well as in 
(\ref{amplfin}) below, are calculated only up to terms proportional to 
$m_n^2$, with this precision they can be taken in the limit $m_n = 0$. 

\begin{figure}[ht]
\centerline{
\epsfig{figure=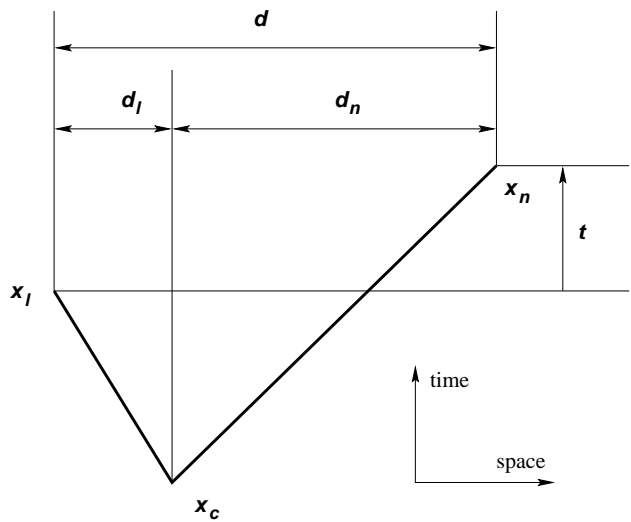,width=0.5\textwidth}}
\caption[]{Time and length definitions in the neutrino source 
rest frame.}
\label{fig}
\end{figure}

 The extremal values of $\lambda_l$ and $\lambda_n$ determined by
the system of equations (\ref{extr1}), (\ref{extr2}) can be easily 
obtained from the kinematics 
of the problem. Let us denote by $t$ and by $d$ correspondingly 
the time difference and the absolute spatial distance between the 
events $x_n$ and $x_l$ in the rest frame of the source [in which 
$p^\alpha = (m, {\bf 0})$], and by $v_l$ and $v_n$ the velocities, 
respectively, of the charged lepton and of the neutrino in this 
frame (see Fig.~\ref{fig}). In the rest frame of the source one has  
\beq
t_{lc} = {d - v_nt \over v_l + v_n} \, , \ \  d_l := \abs{\x_{lc}} = 
v_l t_{lc} \, ,  \ \ \  t_{nc} = {d + v_lt \over v_l + v_n} \, , 
\ \ \ d_n := \abs{\x_{nc}} = v_n t_{nc} \label{xsc} \, , 
\eeq
where also $d_l$ denotes the spatial distance in the source rest 
frame between the point $x_l$ of charged lepton detection and the 
extremal 
point $x_c$, and $d_n$ has the same meaning for neutrino. Then   
\beq
x_{lc}^2 = \left({d - v_n t \over v_l + v_n}\right)^2 \left(1 - 
v_l^2\right) \, , \ \ \ 
x_{nc}^2 = \left({d + v_l t \over v_l + v_n}\right)^2 \left(1 - 
v_n^2\right) \,  , \label{xs}
\eeq
and, using (\ref{extr2}), we obtain 
\beq
\lambda_l = {m_l \over \sqrt{x_{lc}^2}} = {v_l E_l \over d_l} \, , 
\ \ \ 
\lambda_n = {m_n \over \sqrt{x_{nc}^2}} = {v_n E_n \over d_n}\, , 
\label{lambdas2}
\eeq
where $E_l$ and $E_n$ are the energies, respectively, of the charged 
lepton and of the neutrino in the rest frame of the source. 
In this notation 
and in the approximation of $m_n = 0$ for the {\em prefactors\/} 
(but not for the phase) 
the amplitude (\ref{amplmid}) will acquire the form
\beq
{e^{- i\, p \cdot x_l} \over md}\, 
O_D\gamma_\alpha p_n^\alpha O_S \left(m_l - 
\gamma_\beta p_l^\beta\right) \sum_j U_{bj}^\dagger U_{ja}\, 
e^{-i\, p_n \cdot x_{nl}} \, . \label{amplfin}
\eeq
By the way, from the expressions (\ref{lambdas2}) it is clear that 
the extremal values of $\lambda$'s remain finite in the limit of 
$m_n \rightarrow 0$, as was stated above.

\bigskip 

 We shall now estimate the applicability limits of the stationary 
phase approximation used. Our approximation will be good when the 
extremal values of 
$\lambda$'s are much larger than their dispersions determined by the 
matrix $\partial^2 \Phi / \partial \lambda_i \partial \lambda_j$. 
Using (\ref{lambdas2}) and (\ref{mat1})--(\ref{mat3}) we obtain 
after straightforward analysis the conditions
\beq
d_l \le d_n \ll m d_l^2\, \ \ \ \mbox{or} \ \ \ d_n \le d_l \ll 
m d_n^2\, ,  \label{val}
\eeq
under which our approximation is valid. They imply also the 
condition 
\beq
m d \gg 1 \, , \label{val2}
\eeq
which is quite reasonable. 

 In the limit of (\ref{ms}) the energy-momenta 
$p_l$ and $p_n$  change relatively very slightly with the neutrino 
mass specie $j$, and, we remember, our prefactors in (\ref{amplfin}) 
were actually calculated in this limit. In this case the space-time 
behaviour of the 
\mbox{$l$-$\nu$} pair detection probability is given by 
\beq
P_{ba}(x_n, x_l) \propto {1 \over d^2} \left[\sum_j  
\abs{U_{bj}^\dagger U_{ja}}^2 + \sum_{j \ne k} 
\abs{U_{bj}^\dagger U_{ja} U_{ak}^\dagger U_{kb}} 
\cos\left(p_{jk} \cdot x_{nl} + \varphi_{jk}^{ab}\right) \right] 
\, , \label{probfin}
\eeq
where $p_j$ and $p_k$ are neutrino four-momenta of mass species, 
correspondingly, $j$ and $k$, 
$p_{jk} = p_j - p_k$, and $\varphi_{jk}^{ab}$ are constant phases
that stem from the product of matrices $U$ and $U^\dagger$. 
The second 
sum in the square brackets of (\ref{probfin}) describes space-time 
oscillations of the probability. 

 From Eq.~(\ref{probfin}) one obtains the oscillation length and 
oscillation time of the probability considered. Using the 
energy-momentum conservation in the source rest frame of reference 
one has 
\beq
p_{jk}^0 = {\Delta_{jk} \over 2 m}\, , \ \ \ 
|\p|_{jk} \equiv |\p_j| - |\p_k| \approx - {\Delta_{jk} 
\over 2v_l m} = - {\Delta_{jk} E_l \over 2v_n m E_n}\, , 
\label{dels}
\eeq
where approximation uses the assumption (\ref{ms}) 
that neutrino has very small mass, and 
\beq
\Delta_{jk} = m_j^2 - m_k^2 \, . \label{delta}
\eeq 
Thus, oscillation length $L_{\rm osc}$ and oscillation time 
$T_{\rm osc}$ of the $(jk)$-component of (\ref{probfin}) 
in this frame of reference are given, respectively, by 
(we use the limit of $v_n = 1$) 
\beq
L_{\rm osc} ={m \over E_l} L \, , \ \ \ T_{\rm osc} =  
{m \over E_n} L \, , \label{osc}
\eeq
where 
\beq
L = {2 E_n \over \abs{\Delta_{jk}}}  \label{l0}
\eeq
is the standard expression. To proceed to any other reference frame 
what one has to do is to transform the four-vector components 
$p_{jk}^\alpha$ obtained in (\ref{dels}) to this new frame. 
The expressions (\ref{osc}) and the relevant expressions in the 
laboratory frame of reference have been obtained in \cite{DMOS}. 

 If one of the particles, charged lepton or neutrino, is not observed, 
then the probability of detecting the other one is uniform in space 
and time. This is quite obvious without any calculations and is due 
to the fact that the $l$-$\nu$ pair creation probability for a 
monochromatic source 
is homogeneous in space and time. If neutrino is not detected, 
oscillations in the charged lepton detection probability disappear also 
due to orthogonality of neutrino mass eigenstates. This last cause will 
operate with any type of source, not necessarily monochromatic. 
Specifically, it is the necessity of summing the probability over 
the neutrino flavour index $b$ that will eliminate the oscillatory 
terms in this case. 
A detailed discussion of these issues is presented in \cite{DMOS}.

\section{Wave-packet source} \label{pack}

 First of all let us analyse in a little more detail the 
{\em effective\/} region of integration over $x_c$ in (\ref{ampl}) in 
the case of monochromatic source considered in the previous section. 
In other words, it is the region of constructive interference, from 
which most of the contribution to the integral in (\ref{ampl}) comes.
The extension of this region in space-time around the extremal point
$x_c$ is determined by the covariance matrix (\ref{var1}) for given 
values of $\lambda$'s, and by the variation of $\lambda$'s that 
are determined by the covariance matrix $\partial^2 \Phi / \partial 
\lambda_i \partial \lambda_j$ with the components 
(\ref{mat1})--(\ref{mat3}). First, using (\ref{var1})
and (\ref{lambdas2}) we obtain the estimate of the linear dimensions 
$\delta x$ of the effective region of integration for fixed extremal 
values of $\lambda$'s as
\beq
\delta x \simeq \left(\lambda_l + \lambda_n\right)^{-1/2} = 
\sqrt{d_l d_n \over E_n d} \le \sqrt{d \over E_n}\, . \label{delx}
\eeq
Next, we must estimate the linear dimensions $\delta x_c$ of 
the spread of the extremal value $x_c(\lambda_l, \lambda_n)$ 
caused by the spread $\delta \lambda$ of the values of 
$\lambda$'s. This latter spread can be estimated using (\ref{det})
as 
\beq
\delta \lambda \simeq \left|\det\left({\partial^2 \Phi \over 
\partial \lambda_i \partial \lambda_j}\right)\right|^{-1/4} \approx
\sqrt{\lambda_l \lambda_n (\lambda_l + \lambda_n) \over 
p_l \cdot \p_n} \, . \label{dell}
\eeq
Then using (\ref{partxc}), (\ref{xsc}), (\ref{lambdas2}) and the 
condition $\delta \lambda \ll \lambda_{l,n}$ provided by 
Eq.~(\ref{val}), we will have the estimate 
\beq
\delta x_c \simeq {d \, \delta \lambda \over \lambda_l + \lambda_n}
\simeq d \, \sqrt{\lambda_l \lambda_n \over p_l \cdot p_n 
(\lambda_l + \lambda_n)} \approx \sqrt{d \over m} \, . \label{delxc}
\eeq
Approximations ``$\approx$'' in (\ref{dell}) and (\ref{delxc}) use 
smallness of the neutrino mass and become equalities in the limit 
$m_n \rightarrow 0$. 
The last value in (\ref{delx}) is larger than that in (\ref{delxc}) 
hence the dimension of the region of constructive interference will 
be estimated by (\ref{delx}). 

 In realistic situations the source of neutrinos can often be 
approximated by a wave packet with sharp energy and momentum 
distribution (therefore small relative energy-momentum spread). Let 
us suppose that in the rest frame of the source the spread in the 
coordinate space is $\sigma_x$, in momentum space $\sigma_p$, so that 
$\sigma_x \sigma_p \sim 1$. The source has also finite coherence time, 
in the case of pion this will be determined by its lifetime, or by its 
collision time with the environment. Typically, however, the time 
spread $\sigma_t$ of the wave packet is much larger than its spatial 
spread and the latter will determine most of the interesting effects. 

 If the source is to a high precision monochromatic so that its 
spatial and temporal spread is sufficiently large, namely, if 
\beq
\delta x \ll \sigma_x, \, \sigma_t\, , \label{sigs}
\eeq
then we can use the expressions from the previous section for the 
detection probability amplitude whenever the effective region of 
integration over $x_c$ (region of constructive interference) 
lies well within the source wave packet. 
Due to Eq.~(\ref{delx}) the conditions (\ref{sigs}) essentially 
imply
\beq
d \ll E_n \times \min \left( \sigma_x^2, \, \sigma_t^2 \right) \, . 
\label{estim}
\eeq
 
 Spatio-temporal oscillations in the mutual detection probability 
can be observed only up to certain relative distances between the 
detection points of charged lepton and neutrino. We are going 
to determine such maximal distances, called {\em coherence lengths\/}
beyond which oscillations cease to occur. The reason for such 
distances to exist is that 
for different neutrino masses $m_j$ and $m_k$ the corresponding 
centres (extremal points) $x_j$ and $x_k$ of the integration region 
in $x_c$ are different. If they become sufficiently separated in 
space-time they may no longer be able to lie simultaneously within 
the wave packet of the source, thus components in the 
probability amplitude that correspond to different neutrino mass
species will not be able to interfere. 

 Consider this effect quantitatively. The shift four-vector 
$x_{jk} = x_j - x_k$ lies in the plane of the four-momenta 
$p_l$ and $p_n$, and 
can be decomposed into components $x_{jk}^{(l)}$ and $x_{jk}^{(n)}$ 
that go, respectively, along the directions of $p_l$ and 
$p_n$. These components can be easily estimated. Using Eq.~(\ref{xsc}) 
we obtain for the time components (approximation uses $v_n \approx 1$ 
and $\Delta v_l, \, \Delta v_n \ll v_l, \, v_n$)
\ber
&{}&t_{jk}^{(l)} \approx {\Delta v_n \over (v_l + v_n)^2}\, (d + v_l t) 
\approx - {\Delta_{jk} E_l \over 2 m E_n^2}\, d_n \, , \label{dellt} 
\\
&{}&t_{jk}^{(n)} \approx {\Delta v_l \over (v_l + v_n)^2}\, (d - v_n t) 
\approx - {\Delta_{jk} m_l^2 \over 2 m^2 E_n^2}\, d_l\, , \label{delnt}
\eer
where $\Delta_{jk}$ is given by (\ref{delta}). The distances $d_l$ and 
$d_n$ then change as 
\beq
\Delta d_l = - \Delta d_n = v_n \, t_{jk}^{(n)} - v_l \,  t_{jk}^{(l)} 
\approx - {\Delta_{jk} m_l^2 \over 2 m^2 E_n^2}\, d_l + 
{\Delta_{jk} \over 2 m E_n}\, d_n \, . \label{deldl}
\eeq
For large enough absolute values of $t_{jk} = t_{jk}^{(l)} +  
t_{jk}^{(n)}$ and/or $\Delta d_l$ the centres $x_j$ and $x_k$ will 
not be able to lie simultaneously within the source wave packet. 
Components in the 
probability amplitude which correspond to different neutrino mass
species will not be able to interfere, and probability will cease to
oscillate. This will determine coherence lengths of the 
detection probability oscillations in various cases.  

 In the case 
\beq
d_n = {m_l^2 \over m E_n}\, d_l  \label{one}
\eeq
from (\ref{deldl}) it follows that $\Delta d_l = 0$, and the shift 
$x_{jk}$ is in the 
temporal direction in the source rest frame. For sufficiently large 
value of $d = d_l + d_n$ the absolute value of the shift 
$t_{jk}$ becomes larger than the extension $\sigma_t$ of the 
source wave packet in the temporal direction. This determines 
coherence length $L_c$ -- the largest value of $d$ at which 
oscillations can be observed. In the case considered (\ref{one}) it 
is determined by using (\ref{dellt}), (\ref{delnt}) and (\ref{one}) 
as  
\beq
L_c = \sigma_t E_n L \left(1 + {m E_n \over m_l^2} \right) \, , 
\label{lc}
\eeq
with $L$ given by (\ref{l0}). The condition (\ref{estim}) necessary 
for our approximation, will imply that the formula (\ref{lc}) is 
valid for 
\beq
\sigma_t,  {\sigma_x^2 \over \sigma_t} \gg L 
\left(1 + {m E_n \over m_l^2} \right) \, . 
\eeq

 Equation (\ref{one}) implies rather special experimental coincidence 
conditions. In a less special case 
\beq
d_n > {m_l^2 \over m E_n}\, d_l\, , 
\eeq
the second term in the 
last expression of (\ref{deldl}) dominates. In this case the coherence 
length will be determined by the condition that either 
$\abs{t_{jk}}$ becomes larger than $\sigma_t$, or/and 
$\abs{\Delta d_l}$ becomes larger than $\sigma_x$. Using (\ref{dellt}) 
and (\ref{deldl}) we obtain in this case 
\beq
L_c = L\, m \times \min \left(\sigma_x, \sigma_t\, {E_n \over E_l} 
\right) \, . \label{lc2}
\eeq
The condition (\ref{estim}) will determine the validity limit of 
(\ref{lc2}) to be [see Eq.~(\ref{osc})]
\beq
 \sigma_x,  {\sigma_t^2 \over \sigma_x} \gg {m \over E_n} L = 
T_{\rm osc}\, ; \ \ \ 
\sigma_t,  {\sigma_x^2 \over \sigma_t} \gg {m \over E_l} L = 
L_{\rm osc}\, . 
\eeq

 The case 
\beq
d_n < {m_l^2 \over m E_n}\, d_l
\eeq
can describe the situation when the momentum of the charged lepton is 
measured to a good accuracy by measuring its time of flight. Its 
analysis is quite similar to that of the previous cases. The 
oscillations in the probability will disappear at the length 
\beq
L_c = {m^2 E_n \over m_l^2} L \times \min (\sigma_x, \sigma_t)\, ,
\eeq
and the analysis is valid as long as the estimate (\ref{estim}) is 
fulfilled, what gives 
\beq
\sigma_x, \sigma_t, {\sigma_x^2 \over \sigma_t}, {\sigma_t^2 \over 
\sigma_x} \gg {m^2 \over m_l^2} L \, . 
\eeq

 If the source wave packet is sufficiently broad in space 
and if one of the particles, charged lepton or neutrino, is not 
observed, the probability of observing the other one will not 
oscillate in space and time. This is because after integrating the 
probability (\ref{probfin}) over one of the variables 
\{$\x_l$, $\x_n$\} the oscillatory 
terms are averaged to approximately zero. However, with the source 
wave packet sufficiently narrow in space (but still such that the 
condition (\ref{sigs}) or its equivalent (\ref{estim}) holds), 
namely,  
\beq
\sigma_x \ll  L\, ,  \label{opp}
\eeq
even if the accompanying
charged lepton is not observed the neutrino flavour oscillations 
can be observed relative to the source spatial position. Indeed, when 
integrating the detection probability $P_{ba}\left(x_n, x_l\right)$ 
over $\x_l$ we notice that only in a small region of $\x_l$ 
the probability will be nonzero and will be given by (\ref{probfin}). 
This will be the region for which the extremal values of $x_c$ lie 
within the wave packet of the source. The linear dimensions of this 
region are similar to $\sigma_x m / E_l$, and Eq.~(\ref{opp}) follows
from the condition that these dimensions be much smaller than the 
oscillation length $L_{\rm osc}$ which is given by (\ref{osc}). 
Then, after integration 
over $\x_l$ the phases of the oscillatory terms in (\ref{probfin}) 
will be fixed and their dependence on the value of $\x_n$ will 
remain. Detection of neutrino alone is the case most frequently 
discussed in the literature. 

 Consider this effect more thoroughly. The phase $p_{jk} 
\cdot x_{nl}$ in the oscillating term in the probability (\ref{probfin}) 
can be written in terms of the distances $d_l$ and $d_n$ introduced in 
(\ref{xsc}). Choosing the $z$-axis in the direction of $\p_n$ in 
the source rest frame one will have 
\beq
p_{jk} = \left(p^0_{jk}, 0,0, |\p|_{jk}\right)
\, , \ \ \ x_{nl} = \left(t, 0, 0, d\right)\, . 
\eeq
Taking into account the values $p^0_{jk} = \Delta_{jk}/2m$,  
$|\p|_{jk} \approx - \Delta_{jk} E_l / 2v_n m E_n$ 
[see Eq.~(\ref{dels})]
and using (\ref{xsc}), one obtains the standard expression for 
the phase 
\beq
p_{jk} \cdot x_{nl} = {\Delta_{jk} \over 2v_n m}\, \left(d_n - 
{E_l \over E_n}\, d_l \right) + {\Delta_{jk} E_l \over 2v_n m E_n}\, 
\left(d_n + d_l \right) = {d_n \over v_n L} = {t_{nc} \over L} 
\, . \label{delp}
\eeq
From this expression it is again clear that if the distance $d_n$ 
is fixed with accuracy better than $L$ by the position of the 
source wave packet relative to the neutrino detector, the phases
of the probability oscillations will remain fixed even after 
integration of (\ref{probfin}) over the unobserved point $\x_l$, 
and the oscillations will be observed with respect to
the value of $d_n$. This condition again leads to the estimate 
(\ref{opp}). 

 Remarkably, the phase (\ref{delp}) does not depend on $d_l$. 
This fact can also be 
explained as follows. If $d_n$ is fixed and $d_l$ is changing, this 
means that the four-vector $x_{nl}$ changes (say, by amount 
$\Delta x_{nl}$) in the direction of the charged lepton's
four-momentum $p_l$ (this is clear from Fig.~\ref{fig}). 
Then the oscillation phase change is  
$p_{jk} \cdot \Delta x_{nl} \propto p_{jk} \cdot p_l = 
- \Delta p_l \cdot p_l = - \Delta \left(p_l^2\right) / 2 = 0$. The 
expression (\ref{delp}) for the phase coincides with that derived in 
\cite{DMOS} (see Eqs.~(47) and (54) of this reference).

 As noted already at the end of the previous section, and as it 
was discussed in \cite{DMOS}, if neutrino 
is not detected oscillations in the charged lepton detection 
probability disappear in any case due to orthogonality of neutrino 
mass eigenstates. Specifically, it is the summation of 
(\ref{probfin}) over the neutrino flavour index $b$ that will 
eliminate the oscillatory terms.

\section{Source in a localised space-time region} \label{local}

 First consider a hypothetical process in which neutrino, together with 
a charged lepton, is created at a fixed space-time point $x_c$. Note 
that in this case the energy and momentum of the neutrino created is 
totally undetermined. 
The probability amplitude of detecting neutrino-induced charged 
lepton production of flavour $b$ at the space-time point $x_n$ will 
contain the factor
\beq
A_{ba}(x_n, x_c) = \sum_j U_{bj}^\dagger U_{ja} \tilde S_j(x_{nc})\, , 
\label{am}
\eeq
if the charged lepton created together with the neutrino at point $x_c$ 
is of flavour $a$.

 Again, as was already discussed in Sec.~\ref{prel}, the total 
amplitude that 
will describe the detection of neutrino event will contain, besides 
the factor (\ref{am}), also those related to the processes of 
creation, propagation and detection of other particles involved. 
These factors, however, are of particular nature, and do not affect 
the dependence of the probability amplitude of neutrino detection 
on the space-time points $x_n$ and $x_c$, hence they will 
be omitted.  

 The propagator $S_m(x)$ of Eq.~(\ref{prop}) has the leading 
asymptotic behaviour (see. e.g., \cite{GR})
\ber 
S_m(x) \sim \left({e^{3\pi i/4} \over 4\sqrt2\pi^{3/2}}\right) 
{m^{3/2} \over (x^2)^{3/4}}\, \left(1 + {\gamma_\alpha x^\alpha 
\over \sqrt{x^2}} \right) 
\exp\left(- i\, m\sqrt{x^2}\right)\, , \ \  &\mbox{for $m\sqrt{x^2} 
\gg 1$}\, , \label{propl} \\
S_m(x) \sim 
{\gamma_\alpha x^\alpha \over 2\pi^2 \left(x^2\right)^2}
\, , \ \  &\mbox{for $m\sqrt{x^2} \ll 1$}\, , \label{props}
\eer 
Hence, oscillations in the neutrino detection probabilities can 
develop in space and time 
around $x_n$ only when $m_j\sqrt{x_{nc}^2} \gsim 1$ at least for 
the largest of the neutrino masses, since $\tilde S_j(x_{nc})$ do 
not differ for different $j$ in the opposite limit $m_j\sqrt{x_{nc}^2}
\ll 1$. Consider, therefore, the case of $m_j\sqrt{x_{nc}^2} \gg 1$ 
for all $j$. In this limit the amplitude (\ref{am}) up to one and the 
same factor will be given by 
\beq
A_{ba}(x_n, x_c) \propto  \sum_j m_j^{3/2} 
U^\dagger_{bj} U_{ja} \exp\left(- i\, m_j\sqrt{x_{nc}^2}\right)\, . 
\label{am1}
\eeq
Note the mass-dependence of the coefficients in the last equation. 
The space-time variation of the probabilities of the corresponding 
processes will be given by 
\ber
&{}&\hskip-1cm P_{ba}(x_n, x_c) = \tr \left[\cdots 
A_{ba}^\dagger(x_n, x_c) 
\cdots A_{ba}(x_n, x_c) \cdots \right] \propto \nonumber \\ 
&{}&\hskip-1cm \sum_j m_j^3 \abs{U^\dagger_{bj} U_{ja}}^2  + 
\sum_{j \ne k} \left(m_j m_k\right)^{3/2} \abs{U^\dagger_{bj} 
U_{ja} U^\dagger_{ak} U_{kb}} \cos\left(m_{jk}
\sqrt{x_{nc}^2} + \varphi_{jk}^{ab} \right)\, . \label{prob}
\eer
where
\beq
m_{jk} = m_j - m_k \, ,
\eeq 
and $\varphi_{jk}^{ab}$ are constant phases that stem from the 
product of 
matrices $U$ and $U^\dagger$. The last term in the equation 
(\ref{prob}) describes space-time oscillations of the 
probabilities.

 As an illustration consider mixing between two mass eigenstates 
$\nu_1$ and $\nu_2$, with two flavour eigenstates $\nu_\mu$ and 
$\nu_e$. Let, for definiteness, $m_1 > m_2$,
\beq 
\nu_\mu = \nu_1\cos\theta + \nu_2\sin\theta\, , \ \ 
\nu_e = - \nu_1\sin\theta + \nu_2\cos\theta\, , 
\eeq
and let neutrino be created in a flavour eigenstate $\nu_\mu$. 
The corresponding probabilities of detecting muon and electron 
events at $x_n$ will be  
\ber
&{}&\hskip-1cm P_\mu(x_n, x_c) 
\propto m_1^3\cos^4\theta + 
m_2^3\sin^4\theta + 2\left(m_1 m_2\right)
^{3/2}\sin^2\theta\cos^2\theta\cos\left(\Delta m 
\sqrt{x_{nc}^2}\right) \, , \label{mu} \\
&{}&\hskip-1cm P_e(x_n, x_c) \propto \left[m_1^3 + 
m_2^3 - 2\left(m_1 m_2\right)^{3/2}\cos\left(\Delta m 
\sqrt{x_{nc}^2}\right) \right]\sin^2\theta\cos^2\theta\, , 
\label{e}
\eer
where $\Delta m = m_1 - m_2$. The last terms in the square brackets 
in the equations (\ref{mu}) 
and (\ref{e}) describe space-time oscillations of the probabilities. 
Due to mass-dependence of the coefficients in these expressions the 
amplitude of oscillations will be suppressed if $m_2 \ll m_1$, and 
$\cos\theta$ is not too small. 

 The conditions $m_j\sqrt{x_{nc}^2} \gg 1$, together with the 
asymptotic form (\ref{propl}) of the propagator, imply that all the 
mass eigenstate neutrinos $\nu_j$ arrive at the space-time point 
$x_n$ practically as on-mass-shell particles, with four-velocity 
$u = x_{nc}/ \sqrt{x_{nc}^2}$. In the opposite limit 
$m\sqrt{x^2} \ll 1$ the Feynman propagator has the asymptotic 
behaviour (\ref{props}) independent of mass. Then, for instance, 
in our example of two neutrinos, in the region 
$m_2\sqrt{x_{nc}^2} \ll 1 \ll m_1\sqrt{x_{nc}^2}$, which exists in 
the case of large relative mass difference, $m_2 \ll m_1$, one 
obtains 
\ber
&{}&\hskip-1cm P_\mu(x_n, x_c) 
\propto \sin^4\theta + \frac\pi8\zeta^3\cos^4\theta  
+ \sqrt{\pi \over 2}\zeta^{3/2}\sin^2\theta\cos^2\theta
\cos\left(\zeta - 3\pi/4\right)\, , \label{mu1} \\ 
&{}&\hskip-1cm P_e(x_n, x_c)  
\propto \left(1 + \frac\pi8\zeta^3 - \sqrt{\pi \over 2}\zeta^{3/2}
\cos\left( \zeta - 3\pi/4\right)\right)\sin^2\theta\cos^2\theta\, , 
\label{e1}
\eer
where 
\beq
\zeta = m_1\sqrt{x_{nc}^2}\, . \label{zeta}
\eeq

 It is important to stress the difference between the cases of 
sufficiently extended wave packet source and the fixed space-time 
point source. In the first case the probability of detecting 
neutrino is given by Eq.~(\ref{probfin}) with the phase 
given by Eq.~(\ref{delp}), in the second case the probability is 
given by Eq.~(\ref{prob}).
The origin of this difference lies in the fact that in the former
case the amplitude (\ref{ampl}) involves integration over $x_c$, 
whereas in the latter case the point $x_c$ is fixed. In the 
asymptotic limit in which Eq.~(\ref{propl}) is valid neutrino 
propagators have strong pre-exponential mass dependence that 
results in the peculiar mass-dependence of the probability 
(\ref{prob}). With propagators in the asymptotic limit 
(\ref{propl}) it can be explicitly demonstrated that integration 
of the amplitude (\ref{ampl}) over $x_c$ in the case of 
monochromatic source produces neutrino-mass-dependent factors 
that cancel out such pre-exponential neutrino-mass-dependence of 
the probability amplitude and also modify the phase of the 
probability amplitude, leading to Eq.~(\ref{amplfin}). 
In view of the derivation of Eq.~(\ref{amplfin}) presented in 
Sec.~\ref{mono} such a demonstration in the main text would 
be redundant. Therefore, in order to make this point clear, we 
perform it in the Appendix.  

 In reality, creation of neutrino cannot occur at a fixed 
space-time point. However, possible creation region might 
happen to be sufficiently localised by the nature of the source 
or by the experimenter, so that the phase differences 
between components of (\ref{am1}) will be well fixed. Thus  
the equations of this 
section will apply to the situation when neutrino creation region 
is localised in space and time in such a way that 
\beq
\delta\left[m_{jk} \sqrt{\left(x_n - x_c\right)^2}\right] 
\lsim 1 \, ,
\eeq 
where by $\delta [f(x_c)]$ we signify characteristic variation of 
$f(x_c)$ due to variation of $x_c$ over the creation 
region. As the probability oscillation phases 
$m_{jk} \sqrt{\left(x_n - x_c\right)^2}$ 
are symmetric with respect to $x_n \leftrightarrow x_c$ interchange, 
this means that the creation region is to be restricted in space and 
time by the {\em oscillation\/} time and length scales in the 
vicinity of the point $x_n$. Since for small variations 
\beq
\delta\left[m_{jk} \sqrt{\left(x_n - x_c\right)^2}\right] \approx
{} - {m_{jk}\, x_{nc} \cdot \delta x_c \over \sqrt{x_{nc}^2}} \, , 
\eeq 
these oscillation 
scales will be determined by the four-momentum differences 
$p_{jk} = m_{jk}\, x_{nc}/\sqrt{x_{nc}^2}$, with fixed 
four-vector $x_{nc}/\sqrt{x_{nc}^2}$ which is the neutrino 
four-velocity $u$ at the detection event $x_n$. 

 Within the model of two neutrinos considered above we will have 
\beq
p_1 - p_2 = u \Delta m = p_1\, {\Delta m \over m_1} \, , 
\eeq
where $p_1$ is the four-momentum of the heavier neutrino mass 
specie at the detection point. In the case of close neutrino 
masses, $\Delta m \ll m_1$,  
the oscillation length and time will be given by 
\beq
L_{\rm osc} \simeq T_{\rm osc} \simeq {m_1 \over E_n \Delta m} 
\approx {m_1^2 \over E_n^2}\, L \, , 
\eeq
where $L$ is given by Eq.~(\ref{l0}) and $E_n$ is the neutrino 
energy at the detection point. Then the equations (\ref{mu}), 
(\ref{e}) will be valid as long as 
\beq
\sigma_x, \ \sigma_t < {m_1^2 \over E_n^2}\, L \, . \label{con1}
\eeq

 In the case of mass hierarchy, $m_1 \gg m_2$ we have 
\beq
p_1 - p_2 \simeq p_1 \, , 
\eeq
so that 
\beq
L_{\rm osc} \simeq T_{\rm osc} \simeq E_1^{-1} \, ,
\eeq
where $E_1$ is the energy of the heavier neutrino at the detection 
point. The equations for the probabilities in this case are valid for
\beq
\sigma_x, \ \sigma_t < E_1^{-1} \, .  \label{con2} 
\eeq

 Due to incoherent distribution of the sources in realistic situations 
(in a supernova or in the Sun) the 
probabilities (\ref{mu})--(\ref{e1}) will be averaged over the sources 
and the oscillatory terms are likely to disappear. Whenever this is 
the case, the probabilities will describe the so-called {\em global\/} 
effects of appearance-disappearance of neutrino flavour species, and 
will look like 
\ber
&{}& P_\mu(x_n, x_c) 
\propto m_1^3\cos^4\theta + 
m_2^3\sin^4\theta \, , \label{mu2} \\
&{}& P_e(x_n, x_c) \propto \left(m_1^3 + m_2^3 \right)
\sin^2\theta\cos^2\theta \, , \label{e2}
\eer
in the case $m_j\sqrt{x_{nc}^2} \gg 1$, and 
\ber
&{}& P_\mu(x_n, x_c) 
\propto \sin^4\theta + \frac\pi8\zeta^3\cos^4\theta \, , 
\label{mu3} \\ &{}& P_e(x_n, x_c)  
\propto \left(1 + \frac\pi8\zeta^3 \right)\sin^2\theta\cos^2\theta
\approx \frac\pi8\zeta^3 \sin^2\theta\cos^2\theta \, , \label{e3}
\eer
in the case $m_2\sqrt{x_{nc}^2} \ll 1 \ll m_1\sqrt{x_{nc}^2}$, where
$\zeta$ is given by (\ref{zeta}). 

 To see whether the assumption of a well-localised neutrino source 
is realistic, let us make some estimates for the cases of solar 
and supernova neutrinos. We take all the data from the book \cite{KP}.
In all these cases one has $\sigma_x \ll \sigma_t$, and it is the value 
of $\sigma_t$ which will be important.\footnote{Note that in \cite{KP} 
as well as in some other literature $\sigma_x$ usually stands for 
the emitted {\em neutrino\/} wave packet spread. In this paper both 
$\sigma_x$ and $\sigma_t$ denote the spread of the neutrino {\em source}.
Also note that we put the speed of light as well as the Planck constant
to unity and measure $\sigma_t$ in units of length.}   
In the case of solar neutrinos 
\beq
\sigma_t \sim 10^{-7}\, \mbox{cm} \, , \ \ \ E_n 
\sim 10 \, \mbox{MeV} \, . 
\eeq
With $\Delta_{12} = m_1^2 - m_2^2 \sim$ \mbox{$10^{-4}$ eV$^2$} 
in the case of close neutrino masses the condition (\ref{con1}),
under which the formulae of this section will apply, will read
\beq
m_1 \gsim 3 \, \mbox{eV}\, . 
\eeq
Since $E_n^{-1} \sim$ \mbox{$10^{- 12}$ cm}, in the case of mass 
hierarchy the formulae of this section will not work.  

 For supernova neutrinos from the core 
\beq
\sigma_t \sim 10^{-14} \div 10^{-13}\, \mbox{cm} \, , 
 \ \ \ E_n \sim 100 \, \mbox{MeV} \, , 
\eeq
so that $E_n^{-1} \sim 10^{-13}$~cm, the condition (\ref{con2}) 
will be on the edge of fulfillment and the formulae of this section 
might be applicable. 

 For supernova neutrinos from the neutrino sphere 
\beq
\sigma_t \sim 10^{-9}\, \mbox{cm}\, , \ \ \ 
E_n \sim 10 \, \mbox{MeV} \, , 
\eeq
in the case of mass hierarchy the expressions of this section will 
not be applicable. 
In the case of close neutrino masses for $\Delta_{12} = m_1^2 - 
m_2^2 \sim$ \mbox{$10^{-4}$ eV$^2$} we will have the condition 
\beq
m_1 \gsim 0.3 \, \mbox{eV} \, , 
\eeq
for which the relevant expressions of this section will apply.

 In connection with the above examples we must note that the 
probabilities (\ref{mu})--(\ref{e1}) will refer to neutrinos as they 
appear from the source. Subsequent neutrino scattering off the 
particles of the solar or supernova media will result in the 
well-known Mikheyev-Smirnov-Wolfenstein effect (see \cite{BV,KP}), 
which is not considered in this paper.

\section{Summary} \label{sum}

 In this paper we treated the problem of neutrino flavour 
oscillations by consistently using space-time description of the 
relevant processes of particle creation and subsequent detection. We 
described the EPR-like experiments of detecting neutrino together 
with the accompanying charged lepton, as well as the standard textbook 
examples of neutrino flavour space-time oscillations, without invoking 
{\em a priori\/} the notion of particle trajectories. From our 
analysis it is also clear why in fact it is possible to use such a
notion. The effective integration region (the region of constructive 
interference) in the probability amplitude (\ref{ampl}) over 
the space-time point $x_c$ of particle creation is localised around 
the place determined by particle classical trajectories, and the 
contribution to the phase of the amplitude comes mainly from the action 
along these trajectories.  In the case of wave packet neutrino source 
our treatment enabled us to obtain in a rather simple way the 
coherence lengths of the oscillations. We also considered the case 
of neutrino source strongly localised in space and time and in this 
case found out dependence of the probability oscillation amplitude 
on the neutrino masses.

\section*{Acknowledgments}

For valuable comments the author is grateful to E.~V.~Gorbar and 
A.~K.~Ray. He also acknowledges kind hospitality and stimulating 
scientific atmosphere of IUCAA where this paper was written. 
This work was supported in part by the Foundation of Fundamental 
Research of the Ministry of Science of Ukraine under the grant 
No 2.5.1/003.

\appendix

\renewcommand{\theequation}{A\arabic{equation}}
\setcounter{equation}{0}

\section*{Appendix: Alternative derivation of the probability 
amplitude.}

 In this Appendix we shall derive the expression for the probability 
amplitude (\ref{amplfin}) in the case of monochromatic source, in the
limit $m_n\sqrt{x_{nc}^2} \gg 1$, $m_l\sqrt{x_{lc}^2} \gg 1$, using 
the asymptotic expression (\ref{propl}) for the propagators 
in the coordinate representation. 
The integral over $x_c$ in (\ref{ampl}) will be evaluated in the 
stationary phase approximation, according to the assumption that the 
main contribution comes from the region of stationary phase of the 
integrand. 
This phase stems from the propagators and from the source current 
in (\ref{ampl}) and is given by the expression 
\beq
\Phi(x_l, x_n, x_c) = {} - m_l \sqrt{x_{lc}^2} - m_n \sqrt{x_{nc}^2} 
- p \cdot x_c\, , \label{aphase}
\eeq
and it's stationary point is determined from the condition 
\beq
{\partial \Phi \over \partial x_c^\alpha} \equiv m_l\hat x_{lc}^\alpha + 
m_n \hat x_{nc}^\alpha - p^\alpha = 0\, , \label{stat}
\eeq
where the notation is used $\hat x = x / \sqrt{x^2}$. Since 
$\hat x_{lc}$ and $\hat x_{nc}$ are just four-velocities, respectively, 
of the charged lepton and the neutrino at their respective detection points, 
the equation (\ref{stat}) expresses the total energy-momentum conservation. 
Note that for different neutrino masses $m_n = m_j$ the value of $x_c$
determined by Eq.~(\ref{stat}) will be different. 
Let $x_j$ be the solution for $x_c$ of Eq.~(\ref{stat}) with $m_c = m_j$. 
For further convenience we make notation $x_l - x_j = 
x_{lj}$, $x_n - x_j = x_{nj}$, $\hat x_{lj} = u_l$, $\hat x_{nj} = u_n$. 
The phase $\Phi$ can be developed in powers of $x = x_c - x_j$ around 
the stationary point $x = 0$ with the result 
\beq
\Phi(x_l, x_n, x_c) = \frac12 C_{\alpha\beta}\, x^\alpha x^\beta + 
{} \ldots\, , \label{dev}
\eeq
where 
\beq
C_{\alpha\beta} = c_l (u_l)_\alpha (u_l)_\beta + c_n (u_n)_\alpha 
(u_n)_\beta - \left(c_l + c_n\right) g_{\alpha\beta}\, , \label{cov}
\eeq
\beq
c_l = {m_l \over \sqrt{x_{lj}^2}}\, , \ \ \ 
c_n = {m_j \over \sqrt{x_{nj}^2}}\, . \label{cees}
\eeq
Dropping the higher order terms in (\ref{dev}), denoted by dots, we will be 
interested in the value of a Gaussian integral
\beq
\int d^4x \exp\left(\frac i2 C_{\alpha\beta}\, x^\alpha x^\beta \right)\, . 
\eeq
This, up to a constant factor, is given by 
$\abs{\det\{C_{\alpha\beta}\}}^{-1/2}$. In terms of the velocities 
$v_l$ and $v_n$, respectively, of the charged lepton and the neutrino 
in the source rest frame the determinant is given by 
\beq
\det\{C_{\alpha\beta}\} = c_l c_n \left(c_l + c_n\right)^2\left[\left(u_l 
\cdot u_n\right)^2 - 1 \right] = c_l c_n \left(c_l + c_n\right)^2\, 
{\left(v_l + v_n\right)^2 \over \left(1 - v_l^2\right) \left(1 - v_n^2
\right)} \, . \label{detc}
\eeq
The values of $c_l$ and $c_n$ given by (\ref{cees}) can be easily 
seen to coincide with the extremal values, respectively, of 
$\lambda_l$ and $\lambda_n$ determined by the system of equations 
(\ref{extr1}), (\ref{extr2}), and given by (\ref{lambdas2}).  

 Combining all the factors in (\ref{ampl}) together we obtain 
the expression for the amplitude in the 
limit of $m_n \rightarrow 0$ for the prefactors and up to an 
irrelevant constant as
\ber 
&{}& \sum_j {c_l c_n \over (c_l + c_n)}\, 
{\sqrt{\left(1 - v_l^2\right)\left(1 - v_n^2\right)} 
\over v_l + v_n}\,  
O_D \gamma_\alpha u_n^\alpha O_S
\left(1 - \gamma_\beta u_l^\beta\right) 
U_{bj}^\dagger U_{ja}\, e^{i \Phi_j} \nonumber \\
&{}&{} = {1 \over m d}  
O_D \gamma_\alpha p_n^\alpha O_S
\left(m_l - \gamma_\beta p_l^\beta\right) \sum_j 
U_{bj}^\dagger U_{ja}\, e^{i \Phi_j}\, , \label{amplres} 
\eer
where $\Phi_j$ is the value of the phase at the stationary point of
$x_c = x_j$ that corresponds to neutrino mass specie $j$. In view 
of the expression (\ref{aphase}) for the phase we see that this value 
is identical to that of Eq.~(\ref{Phistar}). Therefore the expression 
(\ref{amplres}) for the amplitude coincides with that of 
Eq.~(\ref{amplfin}). 

 Note that the pre-exponential factors in the final expression 
(\ref{amplres}) for the amplitude remain finite in the limit of 
$m_n = 0$ in spite of the fact that neutrino propagators 
(\ref{propl}) have pre-exponential factors $m_n^{3/2}$. The reason 
is that the strong neutrino-mass-dependence of these factors has been 
counterbalanced by the neutrino-mass-dependence of the value 
$x_{nj}^2$ as well as of the factor 
$\abs{\det\{C_{\alpha\beta}\}}^{-1/2}$, 
with the determinant given by (\ref{detc}). In the case of fixed 
neutrino creation point there is no integration over $x_c$ and 
factors $m_n^{3/2}$ remain in the probability amplitude, 
Eq.~(\ref{am1}).

\end{document}